\documentclass[conference]{IEEEtran}
\IEEEoverridecommandlockouts
\usepackage{cite}
\usepackage{amsmath,amssymb,amsfonts}
\usepackage{algorithmic}
\usepackage{algorithm}
\usepackage{graphicx}
\usepackage{textcomp}
\usepackage{xcolor}
\usepackage{graphicx}
\usepackage{textcomp}
\usepackage{theorem}
\usepackage{silence}
\usepackage{bm}
\usepackage{subcaption}
\newtheorem{theorem}{Theorem}
  
\def\BibTeX{{\rm B\kern-.05em{\sc i\kern-.025em b}\kern-.08em
    T\kern-.1667em\lower.7ex\hbox{E}\kern-.125emX}}
\usepackage[hidelinks]{hyperref}
\hypersetup{
    colorlinks=false,   
    linktoc=all,        
    linkcolor= green,     
}
\begin{document}

\title{Movable Antenna Enhanced Downlink Multi-User Integrated Sensing and Communication System\\
}
\author{Yanze Han, Min Li, Xingyu Zhao, Ming-Min Zhao and Min-Jian Zhao \thanks{The authors are with the College of Information Science and Electronic Engineering and also with Zhejiang Provincial Key Laboratory of Multi-Modal Communication Networks and Intelligent Information Processing, Zhejiang University, Hangzhou 310027, China (e-mail: \{YanzeHan,~min.li, zhaoxingyu23,  zmmblack, mjzhao\}@zju.edu.cn).}}

\maketitle

\begin{abstract}
This work investigates the potential of exploiting movable antennas (MAs) to enhance the performance of a multi-user downlink integrated sensing and communication (ISAC) system. Specifically, we formulate an optimization problem to maximize the transmit beampattern gain for sensing while simultaneously meeting each user's communication requirement by jointly optimizing antenna positions and beamforming design. The problem formulated is highly non-convex and involves multivariate-coupled constraints. To address these challenges, we introduce a series of auxiliary random variables and transform the original problem into an augmented Lagrangian problem. A double-loop algorithm based on a penalty dual decomposition framework is then developed to solve the problem.
Numerical results validate the effectiveness of the proposed design,  demonstrating its superiority over MA designs based on successive convex approximation optimization and other baseline approaches in ISAC systems. The results also highlight the advantages of MAs in achieving better sensing performance and improved beam control, especially for sparse arrays with large apertures. 
\end{abstract}


\section{Introduction}
Integrated sensing and communication (ISAC) is emerging as a crucial technology for future sixth-generation (6G) networks, where sensing and communication share the same hardware and frequency band \cite{survey_ISAC}. Various studies have focused on optimizing waveform design, sensing-assisted communication, and communication-assisted sensing to balance performance tradeoffs between communication and sensing\cite{ISAC-funda-xiong}   \cite{ISAC-funda-min}. Among them, multi-antenna and beamforming techniques have been explored to enable flexible integration of sensing and communication signals in the spatial domain, offering enhanced performance compared to separate designs \cite{CRB_opt_liufan}\cite{huang_2023_beampattern}. However, most of these studies assume conventional array architectures with fixed half-wavelength spacing between adjacent antenna elements. This arrangement limits sensing resolution by the number of antenna arrays, and deploying large-scale antenna arrays for further improvement significantly increases costs.

Movable antennas (MAs) have recently emerged as a promising means to enhance Multiple-Input Multiple-Output (MIMO) communication performance. 
By leveraging spatial diversity through adjustable antenna geometry, MAs can achieve superior performance with fewer antennas.  For instance, reference \cite{Wenyuan2023MIMOMAcap} demonstrated MA-based MIMO systems outperform traditional fixed arrays in terms of channel capacity for
point-to-point MIMO systems.  Reference \cite{statistic_csi} further confirmed the potential of MAs to improve the average achievable rate even when only statistical channel state information (CSI) is  available. 
Additionally, reference \cite{PSO} explored multi-user scenarios, showing improved uplink rates by optimizing antenna positions with closed-form beamforming design. 

The aforementioned works have provided many valuable insights into MA-assisted communication. Meanwhile,  in wireless sensing, sparse arrays have also been proven beneficial for increasing the aperture with limited number of antennas, thereby improving detection and estimation performance. Nevertheless, with fixed deployment as usual, it cannot always adapt to the changing environment with suitable geometry. Therefore, recent research has explored the potential of general MAs to enhance wireless sensing, identifying  optimal distributions under various scenarios at the receiver \cite{CRB_MA_sensing}.

Despite their potential, the advantages of employing MAs to balance the tradeoff between sensing and communication in ISAC systems remain insufficiently explored. Several preliminary studies have attempted to address this gap. For instance, in the single-user scenario, reference  \cite{zhouFluidAntennaAssistedISAC2024}    jointly optimized the transmit beamforming and the positioning of fluid antennas—a concept similar to MAs—at both the base station (BS) and the user. Their objective was to maximize the downlink communication rate while meeting constraints on the sensing beampattern gain and transmit power of the BS. Extending this investigation to a multi-user scenario, reference   \cite{zouShiftingISACTradeOff2024c}   jointly optimized the transmit beamforming and port selection of fluid antennas to minimize the total transmit power, ensuring compliance with both communication and sensing requirements. Reference \cite{qinCramerRaoBoundMinimization2024d} instead focused on MA-assisted designs for sensing and communication receivers, and proposed minimizing the Cramér-Rao bound (CRB) for sensing performance at the BS while maintaining a minimum signal-to-interference-plus-noise ratio (SINR) constraint at each user, employing successive convex approximation techniques.

In this paper, we further investigate a MA-assisted downlink multiuser ISAC system, aiming to jointly optimize transmit antenna positions and beamforming design to enhance sensing performance while meeting the communication requirement at each user.  Since the formulated optimization problem is highly non-convex and involves multivariate-coupled constraints, we explore the structural properties of the problem and introduce auxiliary variables to relax the coupling constraints. This allows us to transform the original problem into an augmented Lagrangian
(AL)  formulation. To solve this problem, we develop a double loop algorithm based on a penalty-dual decomposition (PDD) approach. For the inner loop, a block coordinate descent (BCD) method incorporating semidefinite relaxation (SDR) and projected gradient descent (PGD) techniques, is proposed to solve the subproblems. For the outer loop, the dual variables are updated using gradient ascent with decaying penalty parameters. Numerical results illustrate the sensing performance gains achieved by MAs over several baselines in the ISAC system and highlight how MAs can optimize transmit beampatterns and improve beam control capability, especially for sparse arrays with large aperture.
\section{System Model and Problem Formulation}
We consider a multi-user downlink ISAC system, where a multi-antenna base station wishes to communicate with $K$ single-antenna users while performing radar sensing towards a potential point target. Specifically, the transmit array of BS consists of $N_\text{t}\ge K$ antenna elements whose positions $\mathbf{t} = [t_1, t_2,..., t_{N_\text{t}}] \in\mathbb{R}^{N_\text{t}\times 1}$ would be optimized to enhance the sensing performance of ISAC system. The sensing is performed based on echo signals reflected from the target, which are received by conventional fixed antenna arrays or MAs, whose optimization are separated from transmit antennas and has been explored in \cite{CRB_MA_sensing}. 

\subsection{Signal Model and Performance Metrics}
\subsubsection{Signal Model}

Consider a slot-wise transmission with $T_\text{s}$ symbols in each slot. Let $\mathbf{s}_n \in \mathbb{C}^{K \times 1}$ be the dual functional signal adopted by the BS in the $n$-th time slot, with  $\mathbb{E}[s_{k,n}^H s_{i,n}] = 0$ and $\mathbb{E} \left[ \mathbf{s}_n\mathbf{s}_{n}^H \right] \approx {T_\text{s}}^{-1} \sum_{n=1}^{T_\text{s}}{\mathbf{s}_n\mathbf{s}_{n}^H} \approx \mathbf{I}_K$. Let $\mathbf{W}_\text{D} = [\mathbf{w}_1, \mathbf{w}_2, ..., \mathbf{w}_K] \in \mathbb{C}^{N_\text{t}\times K}$ denote the beamforming matrix that satisfies a total power constraint as $\sum_{k=1}^K{||\mathbf{w}_k||^2}\le P_\text{t}$.  Then the transmitted signal by the BS in the $n$-th slot is given by
\begin{equation}
    \label{signal}
    \mathbf{X}_n=\mathbf{W}_\text{D}\mathbf{s}_n. 
\end{equation}
The signal received by the $k$-th user in the $n$-th time slot is 
\begin{equation}
	\label{received signal model}y_{k,n}=\underset{\text{desired}\,\,\text{signal}}{\underbrace{\mathbf{h}_{k}^{H}\mathbf{w}_ks_{k,n}}}+\underset{\text{inter-user}\,\,\text{interference}}{\underbrace{\sum_{i=1, i\ne k }^K{\mathbf{h}_{k}^{H}\mathbf{w}_i}s_{i,n}}}+z_{k,n},
\end{equation}
where $z_{k,n}\sim \mathcal{CN}(0,\sigma_\text{c}^2)$ is the additive Gaussian noise with zero mean and variance $\sigma_\text{c}^2$, and $\mathbf{h}_k\in \mathbb{C} ^{N_\text{t} \times 1}$ is the channel between the BS and the $k$-th user defined as \cite{CS_CE_zhangrui
}
\begin{equation}
    \begin{aligned}
        \mathbf{h}_k = \sum_{l=1}^{L_k} \sigma_{k,l}\mathbf{g}{(\theta_{k,l},\mathbf{t})}, k=1,2,...,K, \label{channel}
    \end{aligned}    
\end{equation}
where $\sigma_{k, l}$ and $\theta_{k, l}$ represent the complex gain and angle of departure (AoD) of the $l$-th path for the $k$-th user, respectively, and  $ \mathbf{g}(\theta_{k,l},\mathbf{t})$ is the steering vector along the direction $\theta_{k,l}$ is defined as 
\begin{equation}
    \mathbf{g}(\theta_{k,l},\mathbf{t})=[e^{-j\frac{2\pi}{\lambda}\text{sin}(\theta_{k,l})t_1},...,e^{-j\frac{2\pi}{\lambda}\text{sin}(\theta_{k,l})t_{N_\text{t}}}]
\end{equation}
with parameter  $\lambda$ being the wavelength.  It is assumed that perfect knowledge about $\mathbf{h}_k$ is available at the BS as in \cite{CS_CE_zhangrui}. 

 \subsubsection{Performance Metrics}
For communication, we utilize the signal-to-interference-plus-noise ratio (SINR) to characterize the quality of service (QoS) for each user. Specifically,  based on the received signal model in \eqref{received signal model}, the SINR at user $k$ is represented as 
\begin{equation}
\label{SINR}
\gamma _k (\mathbf{t},\mathbf{W}_\text{D}) \ =\ \frac{|\mathbf{h}_{k}^H\mathbf{w}_k|^2}{\sum_{i=1, i\ne k }^{K}{|\mathbf{h}_{i}^H\mathbf{w}_i|^2+\sigma _\text{c}^{2}}}\,  .
\end{equation}
\begin{figure}[t]
	\centering
	\includegraphics[width=0.42\textwidth]{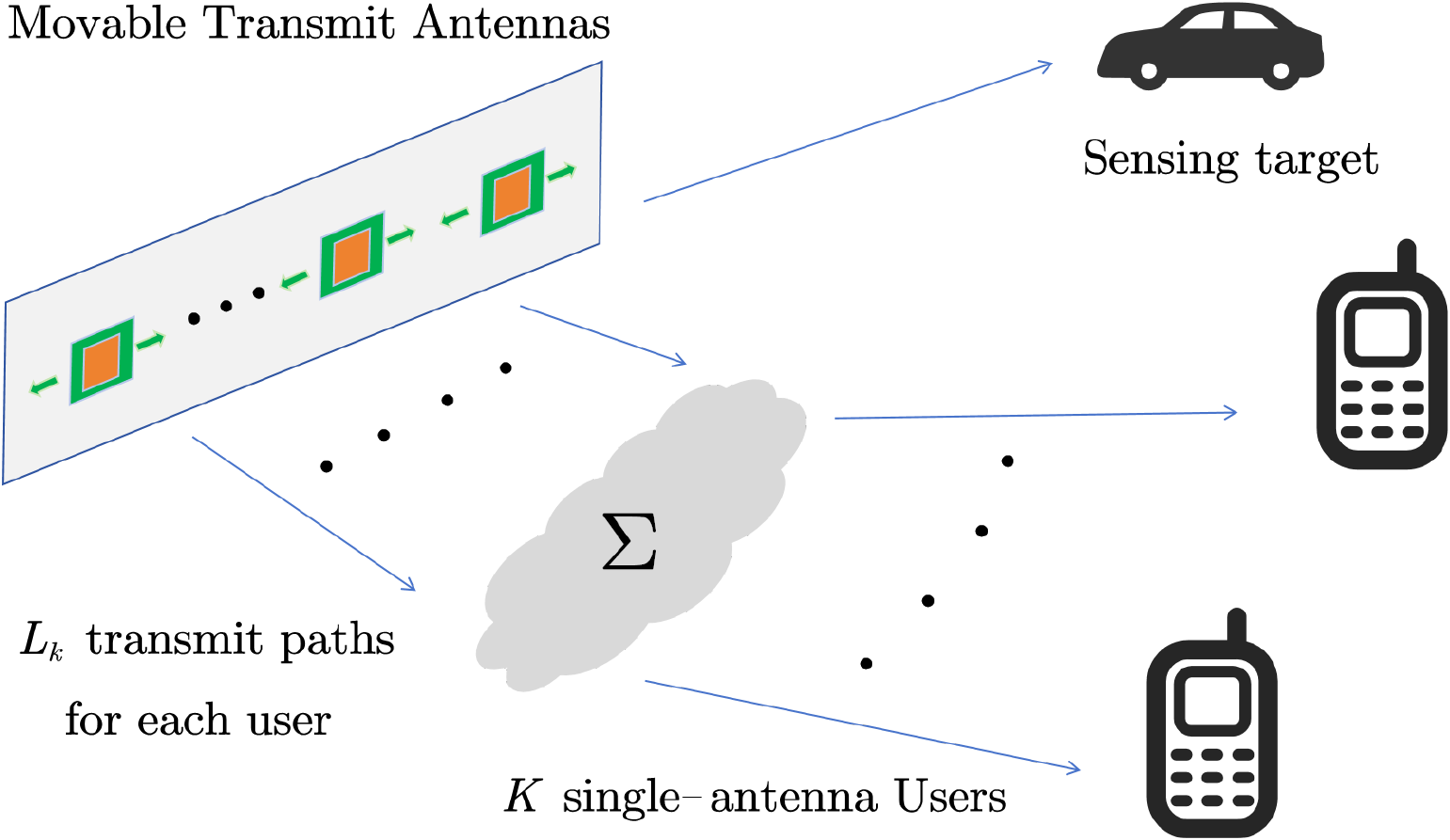}
	\caption{An illustration of the system model considered.}\label{sys_model}
\end{figure}

For sensing, the transmit beampattern that prescribes the distribution of transmit signal power in the angular domain is adopted as the key performance metric \cite{prob}. Specifically, considering the orthogonal data stream assumption in \eqref{signal}, the transmit beam pattern gain along the angle of the target $\theta_\text{s}$ defined as 
\begin{equation}
\begin{aligned}
\label{obj-t}
 P_s  =& \mathbb{E} \left[ \mathrm{tr}\left( \mathbf{a}^H(\mathbf{t})\mathbf{W}_\text{D}\mathbf{s}_n\mathbf{s}_n^H\mathbf{W}_\text{D}^H\mathbf{a}(\mathbf{t}) \right) \right] \\
 \approx& \|\mathbf{a}^H(\mathbf{t})\mathbf{W}_D\|^2 
\end{aligned}
\end{equation}
should be maximized for enhanced sensing performance as shown in 
 \cite{huang_2023_beampattern},  where  $\mathbf{a}(\mathbf{t})= \mathbf{g}(\theta_\text{s},\mathbf{t})$ is the transmit steering vector of MAs on the direction of the sensing target.

\subsection{Problem Formulation }
Given the above signal modeling and performance metrics, we aim to jointly optimize the antenna positions and transmit beam patterns to enhance the sensing performance while ensuring that each user meets a minimum SINR target. Specifically, the problem is formulated as:
\begin{subequations}
    \begin{align}
    (\mathbf{P1}):\underset{\mathbf{t},\{\mathbf{w}_k\}_{k=1}^{K}}{\max}&  \|\mathbf{a}^H(\mathbf{t})\mathbf{W}_\text{D}\|^2  \label{P1} \\
    \mathrm{s}.\mathrm{t}. \,\,\,\,&\gamma_k(\mathbf{t},\mathbf{W}_\text{D}) \ge \Gamma _k, \forall k = 1,2,..., K,\label{P1-C} 
    \\
    &\sum_{k=1}^K{\|\mathbf{w}_k\|^2}\le P_\text{t}, \label{P1-1} 
    \\ 
    & 0 \le t_p \le L, \,\,\forall p=1,2,..., N_\text{t}, \label{P1-2}  \\
    &t_{p+1} - t_p \ge \lambda/2, \,\,\forall p=1,2,..., N_\text{t}-1, \label{P1-3} 
    \end{align}
\end{subequations}
where constraint  \eqref{P1-C} ensures that the communication QoS requirement is met for each user, \eqref{P1-1} corresponds to the transmit power constraint, and  \eqref{P1-2}, \eqref{P1-3} restrict the MAs to a given aperture  $[0,L]$ with minimum spacing $\lambda/2$ between adjacent antenna elements to  prevent the coupling effect \cite{Wenyuan2023MIMOMAcap}.

It is remarked that the problem (P1) is highly non-convex with tightly coupled variables in both  the objective function \eqref{P1} and constraints \eqref{P1-C},  which pose significant challenges on solving the problem. To tackle these issues, 
we introduce a series of auxiliary random variables, transform the problem into a more tractable form and develop an efficient algorithm based on the PDD technique\cite{shi2020penalty}.

\section{Proposed Algorithm}

\subsection{Problem Reformulation} \label{reformulation}
In order to decouple the variables in  \eqref{P1-C} for the users,  we define
\begin{equation}
\label{auxiliary}
    V_{ki}= |\mathbf{h}_{k}^{H}(\mathbf{t})\mathbf{w}_i|^2, \forall k,i=1,2,...,K.
\end{equation}
Moreover, we introduce another set of auxiliary random variables $Q_{k,i}$, such that $Q_{k,i} = V_{k,i}, \forall i, k=1,2,...,K$. In this way, the original problem (P1) is equivalently transformed into 
\parbox{0.488\textwidth} {
\begin{small}
\begin{subequations}

\label{P1_convert}
\begin{align} 
   (\mathbf{P2}): \underset{\{\mathbf{w}_k\}_{k=1}^{K},\mathbf{t},\mathbf{Q}}{\max}  &\|\mathbf{a}^H(\mathbf{t})\mathbf{W}_\text{D}\|^2   	\\
    \,\, \,\,\,\,\,\,\,\,\,\,\,\,\,\,\,\,\,\,\,\,\mathrm{s}.\mathrm{t}.\,\,\,\,\,\,  &\mathbf{Q}=\mathbf{V}(\mathbf{t},\mathbf{W}_\text{D}), \label{equal_constr}
    \\
 & Q_{kk}-\Gamma_k \sum_{i=1, i\ne k}^K {Q_{ki}}-\sigma _{c}^{2}\Gamma _k\ge 0,   \notag\\
 &  \forall k=1,2,...,K\label{eq-SDR-1},   
 \\
 & \eqref{P1-1},\eqref{P1-2} ,\eqref{P1-3}, \notag
 \end{align} 
\end{subequations}
\end{small}}
where $\mathbf{V}$ and $\mathbf{Q}$ are the matrices that are formed by elements $\{V_{k,i}\}$ and $\{Q_{k,i}\}$, respectively. 

It can be seen that the terms in the original \eqref{P1-C}  have been properly decoupled. However, one additional equality constraint \eqref{equal_constr} is introduced, which requires further processing. To address this issue, we relax this constraint and form an AL problem as follows: 

\parbox{0.470\textwidth} {
\begin{small}
\begin{equation}
\label{eq-SDR}
    \begin{aligned}
    (\mathbf{P3}):\underset{\{\mathbf{w}_k\}_{k=1}^K, \mathbf{t}, \mathbf{Q}}{\max}&
    \|\mathbf{a}^H(\mathbf{t})\mathbf{W}_\text{D}\|^2  -\frac{1}{2\rho}{\|\mathbf{Q}-\mathbf{V}(\mathbf{t}, \mathbf{W}_\text{D})+\rho \boldsymbol{\xi} \|^2}& \\
    \mathrm{s}.\mathrm{t}.&\,\, \eqref{P1-1},\eqref{P1-2},\eqref{P1-3} ,\eqref{eq-SDR-1}.
    \end{aligned}
\end{equation}
\end{small}}
where $\boldsymbol{\xi}\in \mathbb{R}^{K \times K}$ denotes the dual variable associated with the equality constraint \eqref{equal_constr}, and $\rho > 0$ denotes the penalty factor. It is noted that as $\rho\to0$,  the constraint violation $||\mathbf{Q}-\mathbf{V}||_\infty$ is enforced to zero, and thus the  equality constraint \eqref{equal_constr} is satisfied. Therefore problem (P3) is equivalent to the original problem asymptotically.

With the above transformations and following the PDD framework in 
\cite{shi2020penalty}, we now propose an iterative algorithm to solve problem (P3). In particular, with the dual variables and penalty factor fixed, the inner-loop problem is solved using the BCD-based method, while with the inner loop variables fixed, the dual variables and penalty factor are updated in the outer loop. In what follows, we will elaborate on how to solve the inner loop problem and discuss the update mechanism of the dual variables and penalty factor.

\subsection{Subproblem With Respect to \texorpdfstring{$\{\{\mathbf{w}_k\}_{k=1}^K,\mathbf{Q}\}$}{}  \label{section1}}
In this subproblem, we first optimize the beamforming vectors $\{\mathbf{w}_k\}_{k=1}^{K}$ and auxiliary matrix $\mathbf{Q}$, with fixed antenna positions $\mathbf{t}$ by utilizing the semidefinite relaxation 
 (SDR) method. Let $\mathbf{W}_k=\mathbf{w}_k\mathbf{w}_k^H$ with $\text{rank}(\mathbf{W}_k)=1$,  we can rewrite the objective function of (P3) as
 
\parbox{0.469\textwidth} { 
\begin{small}
\begin{equation}
        F(\mathbf{W},\mathbf{Q},\mathbf{t})= -\mathbf{a}^H\left(\mathbf{t}\right) \sum_{k=1}^K{\mathbf{W}_k}\mathbf{a}(\mathbf{t})  +
    \frac{1}{2\rho}{\|\mathbf{Q}-\mathbf{V}+\rho \mathbf{\xi} \|^2},
\end{equation}
\end{small}}
and transform (P3) into the following form: 

\parbox{0.469\textwidth} { 
\begin{small}
\begin{subequations}
\label{subproblem1}
    \begin{align}
    (\mathbf{P3.a}):
\underset{\{\mathbf{W}_k\}_{k=1}^K, \mathbf{Q}}{\min} &\,\, F(\mathbf{W}, \mathbf{Q}, \mathbf{t})   \\
\mathrm{s}.\mathrm{t}. &\sum_{k=1}^K \text{tr}(\mathbf{W}_k)\le P_{\text{t}}, 
\\
&\mathbf{W}_k \succeq 0, \forall k = 1,2,..., K,
\\
&\text{rank}(\mathbf{W}_k)=1,\forall k = 1,2,..,K,  \label{rank}
\\
&\eqref{eq-SDR-1}. \notag 
    \end{align}

\end{subequations}
\end{small}}
It is readily seen that, without the rank-one constraint \eqref{rank},  the optimization problem (P3.a) is a standard semidefinite program (SDP) problem, which can be solved efficiently by off-the-shelf numerical solvers, such as mosek/CVX \cite{boyd2004convex}. In fact, we can further prove that if the problem (P3.a) without constraint \eqref{rank} is feasible,  the optimal solution automatically satisfies the rank-one constraint, as stated in  \textbf{Theorem 1} below. 

\begin{theorem}
    \label{theorem} 
 If the relaxed version of (P3.a)  without constraint \eqref{rank} is feasible, the optimal solution $\mathbf{W}^{*}_k$ of relaxed (P3.a)  always satisfies rank($\mathbf{W}_k^{*}$)=1, $\forall k \in \{1: K\}$. 
\end{theorem}
\begin{IEEEproof} 
    Refer to Appendix A for the detailed proof. 
\end{IEEEproof}

Thus, the vector $\mathbf{w}_k^{*}$ could be recovered by performing eigenvector decomposition on $\mathbf{W}_k^{*}$, and the optimal solution could be therefore found for the subproblem (P3.a). 

\subsection{Subproblem With Respect to \texorpdfstring{\{$\mathbf{t}$\}}{}}
By fixing the beamforming vectors $\{\mathbf{w}_k\}_{k=1}^{K}$ and auxiliary matrix $ \mathbf{Q}$, the subproblem with respect to $\mathbf{t}$ is given by 

\parbox{0.469\textwidth} { 
\begin{small}
    \begin{align}
    (\mathbf{P3.b}):
    \underset{\mathbf{t}}{\min} &\,\,  F(\mathbf{W},\mathbf{Q},\mathbf{t})  \label{subproblem2}\\
\mathrm{s}.\mathrm{t}&.\,\, \eqref{P1-2}, \eqref{P1-3}\notag.
    \end{align}
    \end{small}}
\par Problem (P3.b) is  nonconvex with variable $\mathbf{t}$ involved in the complex exponential term of the objective function, making it difficult to find the optimum solution $\mathbf{t}^{*}$ within the feasible region.   However, due to the problem reformulations above from (P1) to (P3), only simple linear constraints remain in  \eqref{P1-2} and \eqref{P1-3}. Consequently, a simple but effective method based on the PGD algorithm is proposed to solve this problem. The key updating operation involved is 
    \begin{align}
    \label{PGD}
    &\mathbf{t}^{d+1}= \text{Proj}_{\mathbf{t}}\{ t^{d}-\gamma^d\nabla F(\mathbf{t}^d)\}, 
    \end{align}
where $\gamma^d$ denotes the step size in the $d$-th PGD iteration, $\nabla F(\mathbf{t}^d)$ denotes the gradient of $F(\mathbf{t})$ with respect to $\mathbf{t}^d$ which could be computed in closed-form via the principles of complex-matrix derivation and chain rules in as given in Appendix B, and the projection operation $\text{Proj}_\mathbf{t}\{\cdot\}$ is defined as follows:

\parbox{0.469\textwidth} { 
\begin{small}
\begin{equation}
\begin{aligned}
    &\text{Proj}_\mathbf{t}\left( t_p \right) =  \\
    &\begin{cases}
	\max \{0,\min\mathrm{[}L-(N_\text{t}-1)\frac{\lambda}{2},t_1]\}, &\mathrm{if}\,\, p=1\\
	\max \{t_{N_\text{t}-1}+\frac{\lambda}{2},\min\mathrm{[}L,t_{N_\text{t}}]\}, &\mathrm{if} \,\,p=N_{\mathrm{t}}\\
	\max \{t_{p-1}+\frac{\lambda}{2},\min\mathrm{[}L-(N_\text{t}-1)\frac{\lambda}{2},t_p]\},&\mathrm{otherwise}.\\
    \end{cases}
\end{aligned}
\end{equation}
\end{small}}

To ensure the convergence of PGD, the step size $\gamma^{d}$ in each iteration $d$ is determined using the backtracking line search technique  \cite{boyd2004convex}, such that the following condition is met:
\begin{equation}
\label{BLS_condition}
    F(\mathbf{t}^{d+1})  \le F(\mathbf{t}^{d})- \gamma^d \|\nabla_{\mathbf{t}^{d}} F(\mathbf{t})|\|_2^2.
\end{equation}

\begin{algorithm}[!t]
    \renewcommand{\algorithmicrequire}{\textbf{Input:}}
    \renewcommand{\algorithmicensure}{\textbf{Output:}}
		\caption{PDD-JAPB algorithm}
  \label{Algorithm}
\begin{algorithmic}[1]  	
\STATE Initialize variable set  $\mathcal{X}^{(0)}=\{\{\mathbf{w}_k\}_{k=1}^K$, $\mathbf{Q}$, $\mathbf{t}\}$, dual variable matrix $\boldsymbol{\xi}^{(0)}$, penalty factor $\rho^{(0)}$ with decay parameter $c_0$, the maximum numbers of iterations $I_\text{in}^\text{max}$, $I_\text{out}^\text{max}$, and the termination thresholds $\delta_\text{in}$, $\delta_\text{out}$.\\ 
\STATE Initialize the outer iteration index $j=0$.
    \REPEAT
        \STATE Initialize the inner iteration index $i=0$. \\
        \REPEAT 
         \STATE Update  $\{\mathbf{w}_k\}_{k=1}^K$ and  $\mathbf{Q}$  based on  (P3.a);
                    \STATE Update $\mathbf{t}$ by the PGD method based on (P3.b);
                    \STATE Update the inner iteration index: $i \leftarrow i+1$;
                    \STATE $\mathcal{X}^{(i)} =  \{\{\mathbf{w}_k\}_{k=1}^K$, $\mathbf{Q}$, $\mathbf{t}\}$;
                    \\
        \UNTIL {$\frac{F(\mathcal{X}^{(i)})-F(\mathcal{X}^{(i-1)})}{F(\mathcal{X}^{(i)})} < \delta_\text{in}$  or $i>I_\text{in}^\text{max}$}.
         \STATE $\mathcal{X}^{(0)} =  \{\{\mathbf{w}_k\}_{k=1}^K$, $\mathbf{Q}$, $\mathbf{t}\}$; \\
  \STATE  Update $\boldsymbol{\xi}^{(j)}$ and $\rho^{(j)}$ based on \eqref{dual_update} and \eqref{penalty_update}, respectively;\\
  \STATE Update the outer iteration index: $j\leftarrow j+1$;
\UNTIL  {$||\mathbf{Q}-\mathbf{V}||_{\infty}<\delta_\text{out}$ or  $j>I_\text{out}^\text{max}$}.
		\end{algorithmic}
	\end{algorithm}
\subsection{Update of Dual Variables and Penalty Factor}
In the outer loop, the dual variables and the penalty factor in the $j$-th outer iteration can be updated by
\label{Outloop_update}
    \begin{align}
    \boldsymbol{\xi}^{(j+1)} &=  \boldsymbol{\xi}^{(j)} + \frac{1}{\rho^{(j)}}(\mathbf{Q}-\mathbf{V}), \label{dual_update}  \\
    \rho^{(j+1)} &= c_0\rho^{(j)}, \label{penalty_update}
    \end{align}
where $0<c_0<1$ denotes a decay factor for $\rho$. Initially, we set a relatively large value for $\rho$, and then we gradually decrease the penalty factor $\rho$ with decay factor $c_0$. The proposed  PDD-based joint antenna positioning and beamforming (PDD-JAPB) algorithm is summarized in Algorithm 1.

\subsection{Convergence and Computational Complexity Analysis}
As for the convergence, each block in the inner loop could descend to an optimal or stationary solution, and thus the BCD algorithm converges. Moreover, with the theoretical convergence analysis of the PDD framework \cite{shi2020penalty}, the whole algorithm is guaranteed to converge to a set of stationary solutions. As for the computational complexity, it mainly depends on problems (P3.a) and (P3.b). Therefore, the overall complexity of the algorithm is $\mathcal{O}(I_\text{out}I_\text{in}[\log(1/\epsilon_\text{sdr})(KN_\text{t}^{3.5}+K^2N_\text{t}^{2.5}+K^3N_\text{t}^{0.5})+K^2N_\text{t}^3+T_\text{bls}N_\text{t}])$, where $T_\text{bls}$ is the number of backtracking-line search.

\section{Numerical Results}
In this section, numerical results are provided to validate the effectiveness of the proposed design.  Specifically, consider an ISAC system operating at carrier frequency $f_0 = 3\text{GHz}$ with $K=4$ users that are randomly and uniformly generated with a distance $d_k\sim[50,150]$m around BS. The channel between the BS and user $k$ is randomly generated according to \eqref{channel} with $L_k = 12, \forall k$, $\theta_{k,l} \sim U(-\frac{\pi}{2},\frac{\pi}{2})$, and $\sigma_{k,l}\sim \mathcal{CN}(0, \sigma d_k^{-\alpha}/L_k)$, where $\sigma = -40\text{dB}$, $\alpha = 2.8$. Additionally, we set $L=15\lambda$, $\sigma_\text{c}^2=-80\text{dBm}$, $P_\text{t}=30\text{dBm}$,  unless otherwise specified, and let  $I_{\text{out}}^\text{max}=30$,  $I_\text{in}^\text{max}=15$, $\rho^{(0)}=1 $, $c_0=0.6$, $\delta_\text{out} $$=$$\delta_\text{in} = 10^{-5}$, when evaluating the PDD-JAPB algorithm. All the results are averaged over 1000 independent channel realizations.

\begin{figure}[t]
    \centering
    \includegraphics[width=0.7\linewidth]{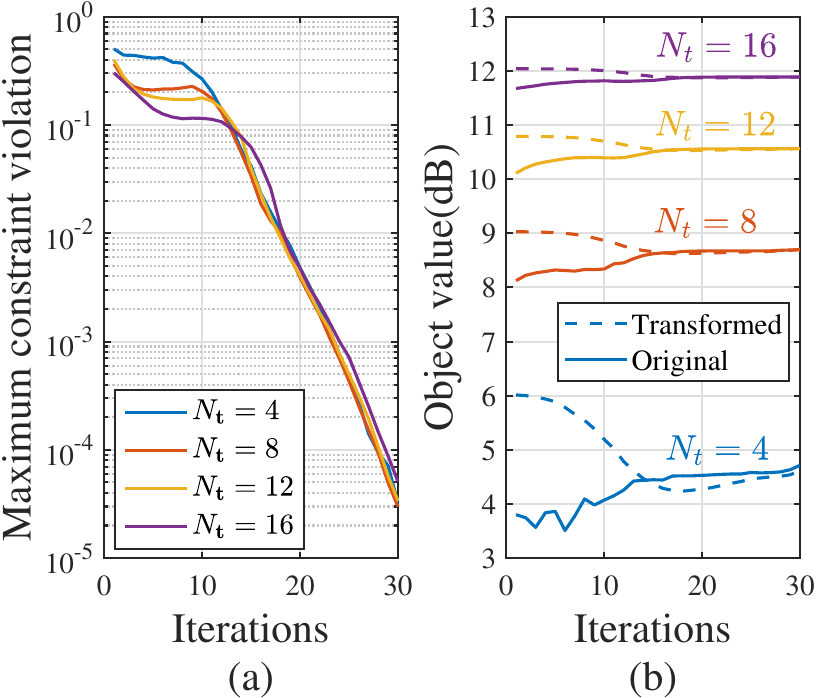}
    \caption{Convergence behavior with $\Gamma_k=10$dB.}
\end{figure}

\par First, we examine the convergence behavior of the proposed algorithm. In particular, with $N_\text{t} \in [4,16]$, Fig. 2(a) shows that the maximum  constraint violation $||\mathbf{Q}-\mathbf{V}||_\infty$  decreases with increasing outer iterations. In addition, Fig. 2(b) further illustrates that the objective value of the transformed problem (P3) approaches to that of (P1)  within tens of outer iterations. 
This confirms the convergence of the proposed algorithm.    

\par Next, we evaluate the performance of the proposed design and compare it against three baseline designs. \textbf{(1) Fixed Antennas (FAs)}: Optimize $\mathbf{W}_\text{D}$ for FAs with adjacent antennas spaced by $\lambda/2$ using SDR.  \textbf{(2) MAs-Random}: Randomly generate the antenna positions $\mathbf{t}$ under constraints \eqref{P1-2} and \eqref{P1-3}, and then optimize $\mathbf{W}_\text{D}$.  \textbf{(3) MAs-SCA}: 
Transform the non-convex objective and SINR constraints into convex forms at each iteration and alternately optimize  $\mathbf{W}_\text{D}$ and $\mathbf{t}$.
Additionally,  we also consider an \textbf{upper bound} on the transmit beampattern gain as a function of $N_\text{t}$, i.e.,   $10\text{log}_{10}(N_\text{t})$ in dB. 

\begin{figure}[t]
    \centering
    \includegraphics[width=0.75\linewidth]{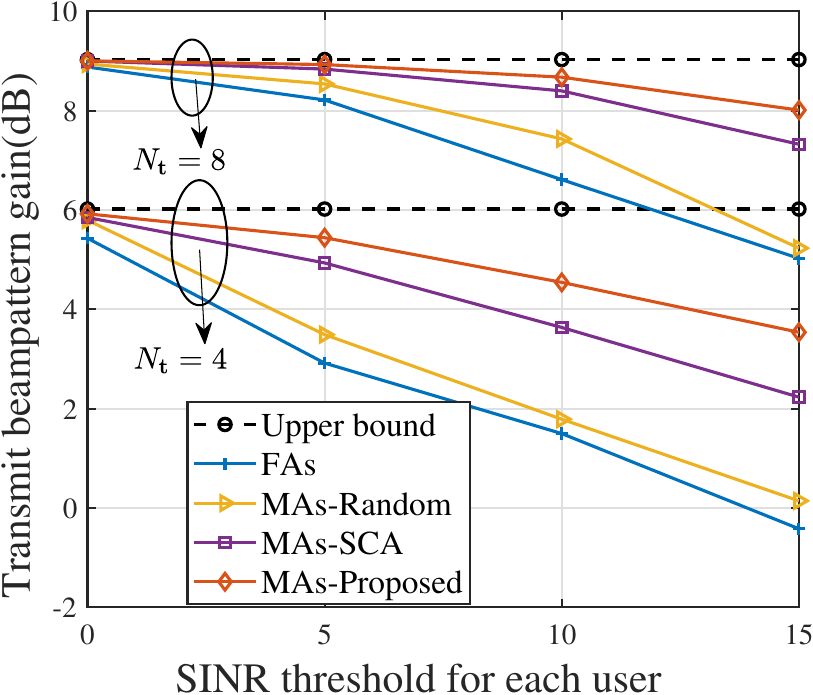}
    \caption{Transmit beampattern gain versus SINR $\Gamma_k$.}
\end{figure}
\par Fig. 3 shows the transmit beampattern gain versus different SINR targets  for both $N_\text{t}= 4$ and $N_\text{t} = 8$.  As expected,  the transmit beampattern gain decreases as the communication SINR target increases for both the MAs-proposed (MAs-PDD-JAPB) and baseline designs. This occurs because the transmit beams must be optimized to meet the more stringent communication requirements, which, in turn, degrades the sensing performance. However, the proposed design is superior to the MAs-SCA,  and significantly outperforms MAs-random and FAs designs, e.g., achieving a gain improvement of over $2$ dB when $N_\text{t} = 4$. This demonstrates the effectiveness of jointly optimizing antenna positions and transmit beampatterns as in the proposed design.

\begin{figure}[t]
    \centering
    \includegraphics[width=0.75\linewidth]{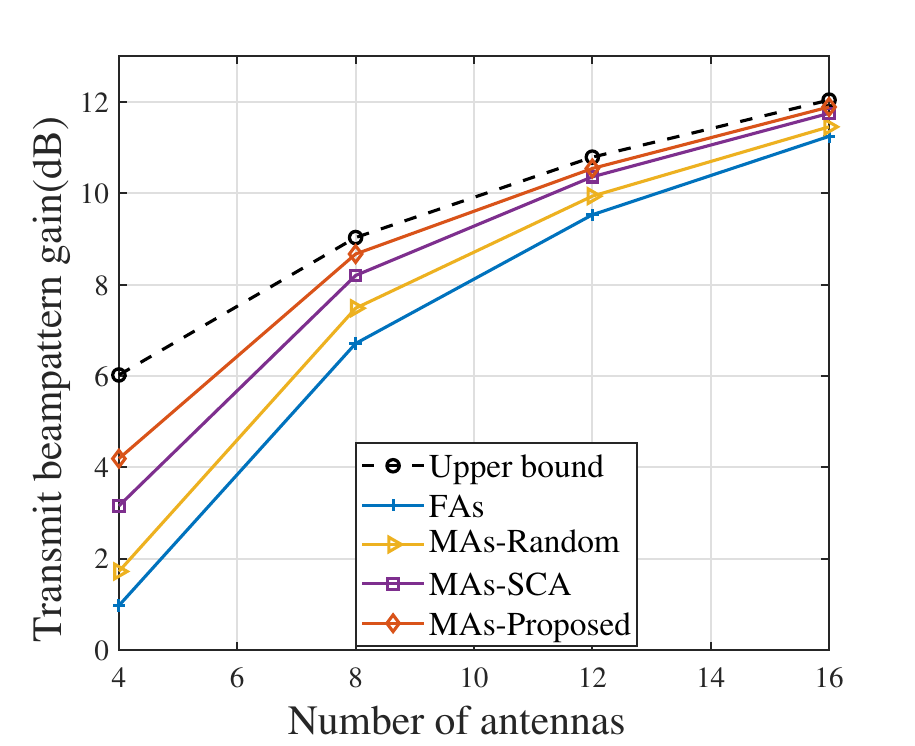}
    \caption{Transmit beampattern gain versus $N_\text{t}$.}
\end{figure}

\par Fig. 4 further evaluates the transmit beampattern gain versus different number of antennas $N_\text{t}$, with the SINR target fixed at 10dB.
Again, the proposed design outperforms all baseline schemes and approaches the upper bound on beampattern gain as $N_\text{t}$ increases. Notably, the performance gap between MAs and FAs narrows when $N_\text{t}$ is large. This is because, with a larger number of antenna elements, the angular resolution of fixed arrays improves significantly, reducing the relative advantage of MAs. 

\par We also consider the CRB for target angle estimation \cite{CRB_opt_liufan} as the sensing performance metric, and evaluate the schemes above using the optimized antenna positions and transmit beamforming configurations that were employed to generate the results in Fig. 4. The corresponding CRB results are presented in Fig. 5. As shown, the MA-assisted designs generally outperform the FAs design. Moreover, the proposed MA optimization consistently surpasses the MAs-SCA approach across various antenna configurations, further demonstrating the effectiveness of the proposed design.

\begin{figure}[t]
    \centering
    \includegraphics[width=0.75\linewidth]{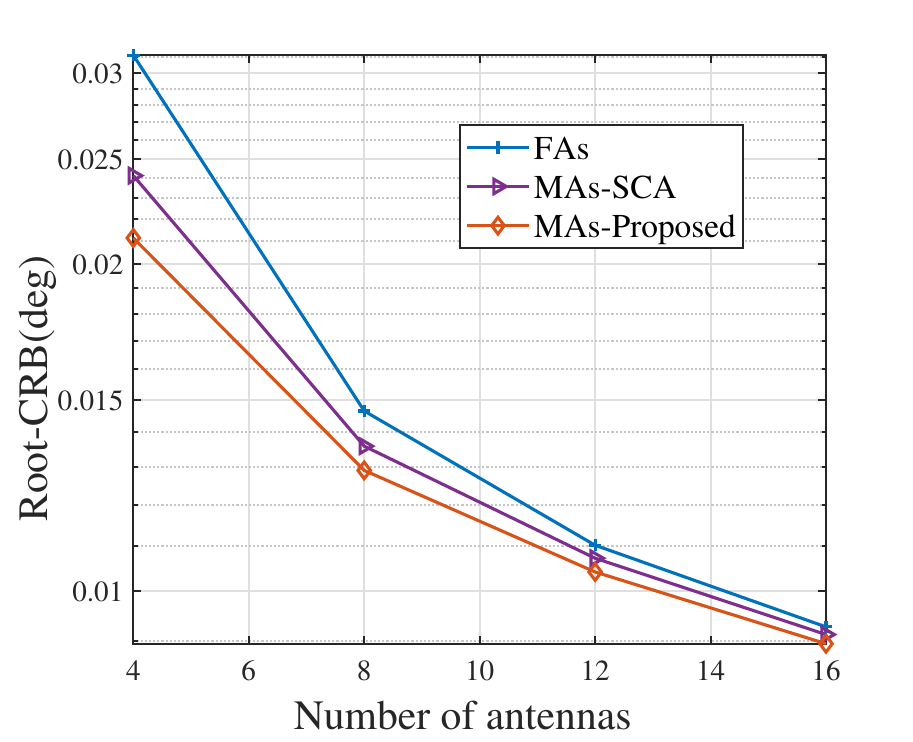}
    \caption{CRB performance evaluation and comparison. }
\end{figure}

\begin{figure}[!t]
    \centering
    \includegraphics[width=0.73\linewidth]{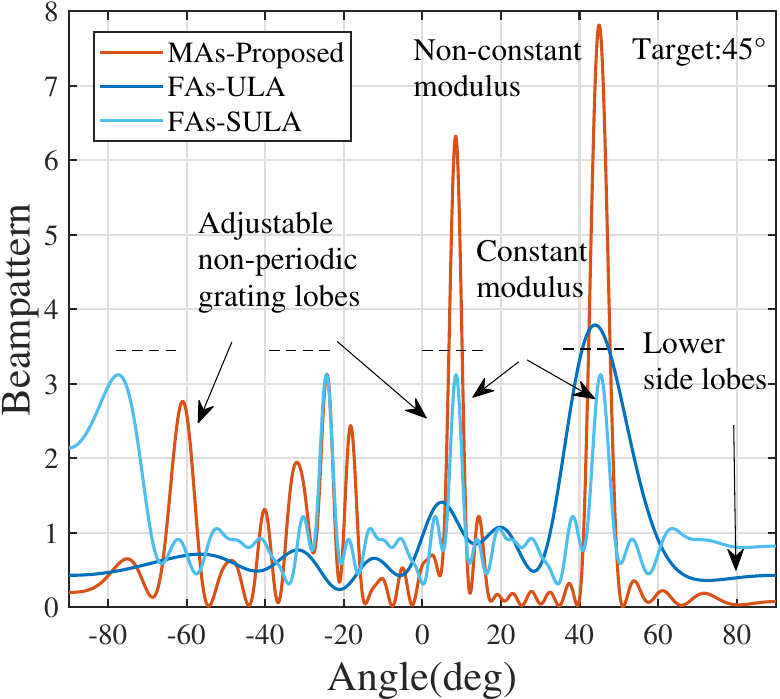}
    \caption{Beampattern comparison with $\Gamma_k=10$dB, $N_\text{t}=8$.}
    \label{fig4}

\end{figure}  
\par Finally, we plot in Fig. 6 an instance of the beampattern attained by MAs and compare it with fixed uniform linear arrays  (ULA) and sparse ULA  (SULA) of the same aperture under $\Gamma_k = 10\text{dB}$ and $N_\text{t}=8$. It can be seen that MA-based sparse arrays generally have narrow beams due to their larger aperture. Additionally, they offer more flexible beam-pointing capabilities to eliminate interference and avoid periodic effects caused by constant modulus grating lobes in the angle domain, which are typical in sparse ULAs.

\section{Conclusion}

In this work, we have investigated the effectiveness of MAs in ISAC systems. We have formulated an optimization problem to maximize the transmit beampattern gain for sensing while meeting users’ QoS requirements. A PDD-based algorithm has been proposed to solve this challenging non-convex and multivariate coupled problem. Simulation results have demonstrated the benefits of MAs in enhancing sensing performance and beam control. Future work could extend this approach to multi-target scenarios and explore other sensing metrics such as the CRB for optimizing the ISAC design.

\section*{Appendix A \label{APA} }
To prove the rank-one property of $\mathbf{W}_k$, we examine the KKT conditions. Let $\{\lambda_k\}_{k=1}^{K+1}\ge 0,\boldsymbol{\Lambda}_k \succeq \mathbf{0}$ denote the dual variables of (P3.a),  then the Lagrangian dual function can be expressed as

\parbox{0.470\textwidth} { 
 \begin{equation}
 \begin{aligned}
     \mathcal{L}_\text{d} =& -\mathbf{a}^H\sum_k^K{\mathbf{W}_k}\mathbf{a} +\frac{1}{2\rho}\|\mathbf{Q}-\mathbf{V}+\rho \mathbf{\xi}\|^2  \\
&-\sum_{k=1}^K{\lambda _k\left(Q_{kk}-\Gamma _k\sum_{i=1\left( i\ne k \right)}^K{Q_{ki}}-\sigma _{c}^{2}\Gamma _k \right)}\,\,  \\
&+\lambda _{k+1}\left(\text{tr}\left( \sum_{k=1}^K{\mathbf{W}_k} \right) -P_\text{t} \right) -\sum_{k=1}^K{\mathbf{\Lambda} _k\mathbf{W}_k}. 
 \end{aligned}
\end{equation}
}
Setting the first-order derivative of $\mathcal{L}_\text{d}$  with respect to  $\mathbf{W}_k$ to be zero yields
\begin{equation}
    \begin{aligned}
        \frac{\partial \mathcal{L_\text{d}}}{\partial \mathbf{W}_k} = &\lambda _{k+1}\mathbf{I}_{N_\text{t}}-\boldsymbol{\Xi}-\mathbf{\Lambda}_k=0,
    \end{aligned}
\end{equation}
where $\mathbf{\Xi}=\mathbf{a}\mathbf{a}^H-\frac{1}{\rho}\sum_{i=1}^K{\left( \tilde{Q}_{ki}-\text{tr}(\mathbf{H}_{i}^{H}\mathbf{W}_k )\right) \mathbf{H}_i}$,  $\mathbf{H}_i=\mathbf{h}_i\mathbf{h}_i^H\in \mathbb{C}^{N_\text{t}\times N_\text{t}}, \forall i\in\{1:K\}$, and $\tilde{\mathbf{Q}}=\mathbf{Q}+\rho\mathbf{\xi}$. Since the dual variable maxtrix $\mathbf{\Lambda}_k$ is positive semi-definite, we obtain 
\begin{equation}
    \begin{aligned}
       \mathbf{\Lambda}_k=\lambda _{k+1}\mathbf{I}_{N_\text{t}} - \boldsymbol{\Xi} \succeq \mathbf{0},
    \end{aligned}
\end{equation}
which implies that  $\lambda_{k+1} \ge \lambda_\text{max}(\boldsymbol{\Xi})$, where $\lambda_\text{max}(\cdot)$ denotes the maximum of the eigenvalue of its argument.  On the other hand, by the complementary slackness condition, we have that 
\begin{align}
    \mathbf{\Lambda}_k\mathbf{W}_k=0, \mathbf{W}_k \ne \mathbf{0},
\end{align}
which implies that $\boldsymbol{\Lambda}_k$ is not full rank. Therefore, it can be obtained that $\lambda_{k+1} = \lambda_\text{max}(\boldsymbol{\Xi})$ and $\text{rank}(\mathbf{\Lambda}_k)=N_\text{t}-1$. Since $\text{rank}(\mathbf{AB})\ge \text{rank}(\mathbf{A})+\text{rank}(\mathbf{B})-N(\text{dim}), $ we thus have that 
\begin{equation}
     \begin{aligned}
         \text{rank}(\mathbf{W}_k)\le  \text{rank}(\mathbf{\Lambda}_k\mathbf{W}_k) - \text{rank}(\mathbf{\Lambda}_k)
         + N_\text{t}=1,
     \end{aligned}
 \end{equation}
which implies that $\text{rank} (\mathbf{W}_k) = 1$.  This completes the proof.

\section*{Appendix B }\label{APB}
Based on the principles of complex-matrix derivation and chain rules, the gradient $\nabla F(\mathbf{t}^x)$ can be calculated as follows 
 
 \parbox{0.470\textwidth} { 
\begin{equation}
    \begin{aligned}
  \label{gradient}
  &\frac{\partial F}{\partial \mathbf{t}}=\frac{\partial \mathbf{a}}{\partial \mathbf{t}}\frac{\partial F}{\partial \mathbf{a}}+\frac{\partial \mathbf{a}^{H}}{\partial \mathbf{t}}\left( \frac{\partial F}{\partial \mathbf{a}^{H}} \right) ^{H}+ \\
  &\sum_{k=1}^K{\sum_{i=1}^K{\frac{\partial F}{\partial B_{ki}}\left( \frac{\partial \mathbf{h}_k}{\partial \mathbf{t}}\frac{\partial B_{ki}}{\partial \mathbf{h}_k}+\frac{\partial \mathbf{h}_{k}^{H}}{\partial \mathbf{t}}\left( \frac{\partial B_{ki}}{\partial \mathbf{h}_{k}^{H}} \right) ^{H} \right)}}.
\end{aligned}
\end{equation}
}
Specifically, define $ \varphi_s = j2\pi\sin\theta_s/\lambda, \varphi_{k,l}=j2\pi\sin\theta_{k,l}/\lambda,B_{ki}=\mathbf{h}_k^{H}\mathbf{W}_i\mathbf{h}_k\in \mathbb{R}^{1\times 1}, A_{ki}=Q_{ki}+\rho \xi_{ki}\in \mathbb{R}^{1\times 1},\mathbf{g}_{k,l}=\mathbf{g}(\theta_{k,l})$, and  $  \mathbf{W}_i=\mathbf{w}_i\mathbf{w}_i \succeq 0 \in \mathbb{C}^{N_{\text{t}}\times N_{\text{t}}}$, then, the terms in \eqref{gradient} can be computed in closed-form as follows:
    \begin{align}
&\frac{\partial F}{\partial \mathbf{a}}=\left(\mathbf{a}^H\mathbf{R}_x \right) ^H,\,\,\frac{\partial F}{\partial \mathbf{a}^H}=\left( \mathbf{R}_x \mathbf{a}\right)^H, \\
 & \frac{\partial B_{ki}}{\partial \mathbf{h}_k}=\,\,\left( \mathbf{h}_{k}^{H}\mathbf{W}_i \right) ^H, \frac{\partial B_{ki}}{\partial \mathbf{h}_{k}^{H}}=\left( \mathbf{W}_i\mathbf{h}_k \right) ^H, \\
 &\frac{\partial \mathbf{a}}{\partial \mathbf{t}}=\mathrm{diag}\left( -\varphi _se^{-\varphi_s \mathbf{t}} \right), 
    \frac{\partial \mathbf{a}^H}{\partial \mathbf{t}}=\mathrm{diag}\left(\varphi_se^{\varphi_s\mathbf{t}} \right),\\
&\frac{\partial F}{\partial B_{ki}}=\,\,2\left( B_{ki}-A_{ki} \right),\frac{\partial \mathbf{h}_k^H}{\partial \mathbf{t}}=\sum_{l=1}^{L_t}{\sigma _{k,l}}\mathrm{diag}\left( \varphi_{k,l}e^{\varphi_{k,l}\mathbf{t}} \right).
    \end{align} 

\bibliographystyle{IEEEtran}
\bibliography{IEEEabrv,ref}

\end{document}